\documentclass{article} 
\usepackage{viva2}
\usepackage{url}
\usepackage{epsfig}
\usepackage{array}
\usepackage{times}

\def\bit{\bfseries}

\let\lcitebracket(
\let\rcitebracket)

\begin{document}

\title{Formal Test Purposes and The Validity of Test Cases\thanks{This paper appeared in the proceedings of the 22nd IFIP WG 6.1 International Conference on Formal Techniques for Networked and Distributed Systems (FORTE 2002), number 2529 Lecture Notes in Computer Science.} }

\author{Peter H. Deussen\\Fraunhofer FOKUS, Berlin, Germany\\ \texttt{deussen@fokus.fhg.de} 
  \and Stephan Tobies\\ Nokia Research Center, Bochum, Germany\\ \texttt{Stephan.Tobies@nokia.com}}

\maketitle

\vspace{-.5cm}

\begin{abstract}
  We give a formalization of the notion of test purpose based on
  (suitably restricted) Message Sequence Charts. We define the validity of
  test cases with respect to such a formal test purpose and provide a simple
  decision procedure for validity.
\end{abstract}

\renewcommand{\epsilon}{\varepsilon}

\newtheorem{algorithm}{Algorithm}{\bfseries}{\itshape}
\newtheorem{lemma}{Lemma}{\bfseries}{\itshape}
\newtheorem{theorem}{Theorem}{\bfseries}{\itshape}
\newtheorem{proof}{Proof}{\itshape}{\rmfamily}

\section{Introduction} 

The quality of a test system directly influences the quality of the
tested implementation: high quality test systems are essential to
obtain high quality implementations.  Hence, a common problem in the
testing area is the so-called ``test the tester''
problem \cite{scheurer98:_revis_compar_autom_gener_manual}: how can the
validity of a test system with respect to a given specification, and
therefore the quality of the test system, be assured? To put it in
conformance testing terminology: how can it be assured that a test
case achieves its test purpose?

One approach used to obtain valid test systems is the derivation of
test cases from formal specifications or test purpose definitions. Other
approaches focus on the manual or automated simulation against a
formal specification (see \cite{schmitt00:_test_gener_autol_testc} for
a description of tools that employ these two approaches).  While many
modern telecommunication protocols come with (semi-) formal
specifications of test purposes, a formal protocol description is
provided only in very few cases (see
\cite{grabowski00:_ttcn_sdl_msc_inap} for a notable exception).  For
example, Internet Protocols defined in RFCs use natural language to
define the semantics of the specification. Due to this, a formal
description of the specification would have to be elaborated to allow
for an automatic generation of valid test cases.  Additionally, even
if formal descriptions are available, automated generation methods
only generate test skeletons that need to be manually refined to
obtain executable test cases for the execution against a concrete system
implementation. For all these reasons, the implementation of test cases
is still performed mainly in a manual manner.

In this paper, we give a new answer to the ``test the tester
problem'', namely, to check the validity of a (possibly hand-written)
test case against a formal test purpose definition. It does not rely
on the existence of a formal description of the system under test
(SUT) or the test system, but requires a formally defined test
purpose. From this test purpose, the allowed and required behavior of
the test case is derived. This information is then used in a guided
simulation of the executable test system to determine whether the test
system is valid with respect to this test purpose. Since our approach
is solely based on test purposes, it is not necessary to develop a
complete formal specification of the system as test purposes are only
a partial description of the system. We use Message Sequence Charts
(MSC) as the formal test purpose description language \cite{ITU-Z120},
which is widely used in the system development process in the
telecommunication area. This allows for an easy re-use of the
uses-cases developed during system design as a solid basis for the
test purpose definition.  This further reduces the work necessary for
the test purpose specification.

Despite the fact that MSCs are widely used to capture test purpose,
theoretical studies of MSCs so far seem to have failed to address the
following issues:

\begin{itemize}
\item What does it mean for a test case to implement a test purpose,
  i.e., when is a test case \emph{valid} w.\,r.\,t.\ a test purpose?
\item When is an MSC a \emph{well-formed} test purpose, i.\,e., when
  does an MSC characterize behavior that is indeed (black-box)
  testable?
\end{itemize}

We address these issues using a semantics for MSC based on pomsets
\cite{pratt86,grabowski81} in the spirit of \cite{kat-lam98}. We then
describe a simple decision procedure for the validity of test cases w.\,r.\,t.\ 
a test purpose and prove its correctness.

The paper is structured as follows: Section \ref{section:msc} of this
paper introduces the partial order semantics of MSCs and their usage
as formal test purposes.  In Section \ref{section:validity}, we define
test case validity, describe the decision procedure and prove its
correctness. Section~\ref{section:implementation} presents one
possible implementation design for an MSC based test validator.
Section \ref{section:outlook} concludes. Proofs of key lemmata and
theorems can be found in the appendix.

\section{Formal Test Purposes}\label{section:msc}

To check (or even define) validity of a test case wrt.\ a test
purposes, we need a formal definition of a test purpose together with
suitable semantics. In this section, we suggest a formalism to formally
express test purposes and establish a set of criteria that guarantee
that a test purposes indeed describes (black-box) testable behavior.

We use Message Sequence Charts (MSCs) to express formal test purposes
because they are widely used to capture test purposes and semantics
based on different approaches are available. We have chosen semantics
based on pomsets~\cite{grabowski81,pratt86} in an adaption of the
definition of \cite{kat-lam98} to better suit our purposes. The
particular choice of semantics of MSCs in not central to our approach,
but obviously some choice has to be made. Using the more operational
semantics from \cite{mauw-reniers97,jonsson01:_execut_seman_msc} would
lead to similar results.

After a short overview on the employed MSC syntax for test purposes,
we recapitulate the pomset-based semantics of MSC and define when an
MSC constitutes as well-formed test purpose.

\paragraph*{Message Sequence Charts.}

\begin{figure}[t]
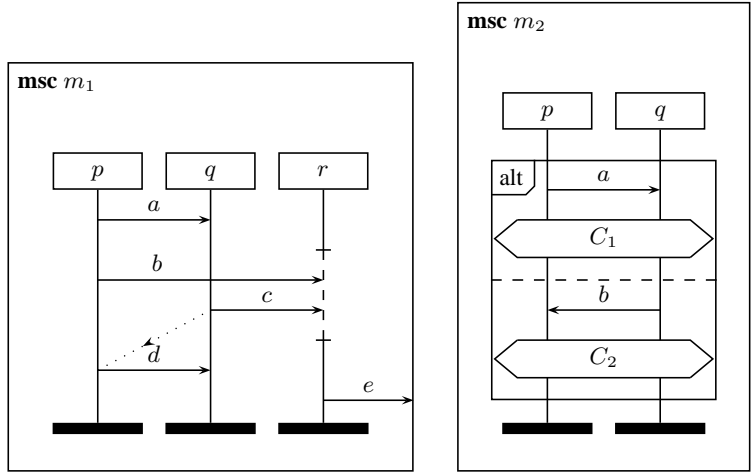

  \begin{center}
    \begin{footnotesize}
      \setmscvalues{small}
         \begin{tabular}{cc}
           \begin{msc}{\ensuremath{m_1}}
             \declinst{p}{}{$p$}
             \declinst{q}{}{$q$}
             \declinst{r}{}{$r$}
             \mess{$a$}{p}{q}\nextlevel
             \coregionstart{r} \nextlevel
               \mess{$b$~~~~~~~~~~~~~~~~~~}{p}{r} \nextlevel
               \mess{$c$}{q}{r} \order{q}{p}[2] \nextlevel
             \coregionend{r}\nextlevel
             \mess{$d$}{p}{q}\nextlevel
             \mess{$e$}{r}{envright}
           \end{msc}
         &
           \begin{msc}{\ensuremath{m_2}}
               \declinst{p}{}{$p$}
               \declinst{q}{}{$q$}
               \inlinestart{alt}{alt}{p}{q} \nextlevel
                   \mess{$a$}{p}{q}\nextlevel
                   \condition{$C_1$}{p,q}\nextlevel[2]
                   \inlineseparator{alt} \nextlevel
                   \mess{$b$}{q}{p}\nextlevel
                   \condition{$C_2$}{p,q}\nextlevel[2]
               \inlineend{alt}
           \end{msc}
         \end{tabular}
     \end{footnotesize}
   \end{center}
   \vspace*{-.5cm}
   \caption{\label{fig:intro-example}Example MSCs.}
   \vspace*{-.2cm}
\end{figure}

The MSCs in Fig. \ref{fig:intro-example} serves as an explanatory
example for the basic MSC language as used throughout this paper. The
most fundamental constructs of MSCs are {\em instances\/} and {\em
  messages}. Instances represent components or communication
interfaces that exhibit a sequential behaviour.  Our example MSC $m_1$
consists of three instances $p$, $q$, and $r$. A message exchange
between a sending instance $p$ and a receiving instance $q$ comprises
two {\em events\/} $\send{p, q}a$ and $\rec{p, q}a$ for sending the
message $a$ at $p$ and for receiving $a$ at $q$, respectively.
Graphically, messages are depicted by arrows between instances
labeled with messages.

Events are considered to be causally or temporally ordered only if
they are located at the same instance (in this case the ordering is
top-to-bottom), or if they are part of the same message exchange.  In
our example $m_1$, the event $\send{p, q}a$ precedes the
events $\rec{p, q}a$ and $\send{p, r}b$, but no assumption on an
ordering of the events $\send{p, r}b$ and $\send{q, r}c$ is expressed,
even if $\send{p, r}b$ is drawn above  $\send{q, r}c$.

There is a way to express the concurrency of events of the same
instance: the concurrent region ({\em coregion}, for short). Coregions
are depicted by dashed sections on the corresponding instance line
bordered by small horizontal bars: the events that occur on this
dashed section are supposed to happen in parallel. In our example,
the events $\rec{p, r}b$ and $\rec{q, r}c$ are temporally unrelated.
On the other hand, it is possible to use {\em general order\/} arrows
(dotted lines between events with an arrow head in their middle
section) to express causal orderings of events on different instances.
In $m_1$, the event $\send{q, r}c$ precedes $\send{p, q}d$.  Finally,
the MSC language allows to express message exchange with the
environment of a MSC; e.\,g. in $m_1$ the message $e$ is send to the
environment of this MSC.

The MSC formalism provides not only communication primitives but also
control structures. For our purposes, only the \emph{alt operator}.
modeling nondeterministic choice, is of importance. $m_2$ in Fig.
\ref{fig:intro-example} shows an example: A choice between sending $a$
from $p$ to $q$ and sending $b$ from $q$ to $p$ is expressed. A final
construct considered in this paper is that of {\em conditions}.
Conditions model global states or predicates related to more than on
instance; $m_2$ contains two conditions $C_1$ and $C_2$. It is not an
easy task to assign a formal meaning to conditions. However, we use
conditions only to express test verdicts and handle them formally in a
special way. We will discuss this topic in detail in a later section.

Other important concepts of the basic MSC language not covered in this
paper are: {\em loop inline expressions\/} (since tests are finite,
loops occurring in test purposes comprises alway finite, fixed
boundaries and therefore can be unfolded), and especially {\em
  timers}, which require extra considerations and will be dealt with
in forthcoming work.

\paragraph*{Expressing test purposes.}

We will use the MSC formalism to capture test
purposes in the following way: the set of instances is partitioned into a non-empty set of
port instances and a non-empty set of SUT instance. Intuitively, the
port instances represent the different ports (PCOs, interfaces) at
which the SUT interacts with its environment. Conditions that span the
port instances are used to assign the test verdicts.

\begin{figure}[t]
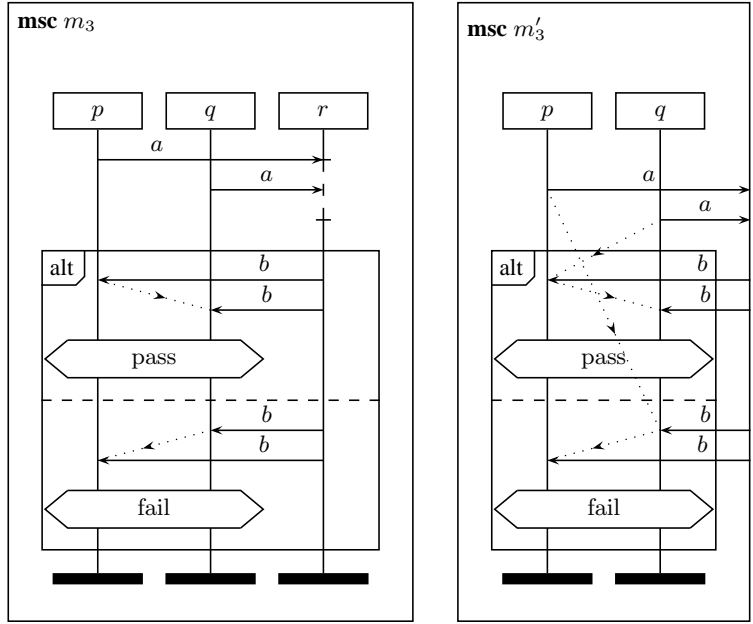

  \begin{center}
    \begin{footnotesize}
      \setmscvalues{small}
       \begin{tabular}{cc}
         \begin{msc}{\protect{\ensuremath{m_3}}}

           \declinst{p}{}{$p$}
           \declinst{q}{}{$q$}
           \declinst{r}{}{$r$}
           \coregionstart{r} 
           \mess{$a$~~~~~~~~~~~~~~~~~~}{p}{r} \nextlevel
           \mess{$a$}{q}{r} \nextlevel
           \coregionend{r}\nextlevel
           \inlinestart{alt}{alt}{p}{r} \nextlevel
           \mess{~~~~~~~~~~~~~~~~~~$b$}{r}{p} \order{p}{q}[1] \nextlevel
           \mess{$b$}{r}{q} \nextlevel
           \condition{$\Vpass$}{p,q} \nextlevel[2]
           \inlineseparator{alt} \nextlevel
           \mess{$b$}{r}{q} \order{q}{p}[1] \nextlevel
           \mess{~~~~~~~~~~~~~~~~~~$b$}{r}{p}  \nextlevel
           \condition{$\Vfail$}{p,q} \nextlevel[2]
           \inlineend{alt}
         \end{msc} &

         \begin{msc}{\ensuremath{m_3'}}
           \nextlevel
           \declinst{p}{}{$p$}
           \declinst{q}{}{$q$}
           \mess{$a$}{p}{envright} \order{p}{q}[8] \nextlevel
           \mess{$a$}{q}{envright}
        \order{q}{p}[2]\nextlevel
        \inlinestart{alt}{alt}{p}{q} \nextlevel
        \mess{~~~~~~~~~~~~~~~~~~$b$}{envright}{p} \order{p}{q}[1]\nextlevel
        \mess{$b$}{envright}{q} \nextlevel
        \condition{$\Vpass$}{p,q} \nextlevel[2]
        \inlineseparator{alt} \nextlevel
        \mess{$b$}{envright}{q} \order{q}{p}[1] \nextlevel
        \mess{~~~~~~~~~~~~~~~~~~$b$}{envright}{p} \nextlevel
        \condition{$\Vfail$}{p,q} \nextlevel[2]
        \inlineend{alt}
      \end{msc}
    \end{tabular}
    \end{footnotesize}
  \end{center}
  \vspace{-.5cm}
  \caption{\label{fig:alternatives}Expressing the same test purposes with and without SUT instances}
  \vspace*{-.2cm}
\end{figure}

The SUT instances are used as ``syntactic sugar'' and serve two
purposes: (1) as communication partners
 for the port instances, and
(2) to impose an ordering of the sequence of messages.  The same could
be achieved by using communication with the environment and
generalized orderings, but our approach leads to a more concise and
intuitive representation of the test purpose and matches the common usage.
Fig. \ref{fig:alternatives} shows the two alternative ways of
depicting a simple test purpose: after having received the message $a$
on both its ports $p$ and $q$ (in arbitrary order), the SUT answers by
sending the message $b$, again both on port $p$ and $q$. If the
message is sent on port $p$ before it is sent on port $q$ then the SUT
shall pass the test, otherwise it shall fail. We will come back to
this example later in this paper.

\subsection{Partial Orders}

We quickly recapitulate how pomsets can be used to assign a semantics
to MSCs. We start by introducing the basic notations used throughout
this paper.

To avoid tedious notation, we fix the following convention: if a
structure $S = \<A, B, \ldots>$ is introduced, the components of $S$
will be denoted by $A_S, B_S, \ldots$

For some set $A$, $\Pow(A)$ is the set of all subsets of $A$.  For $R
\subseteq A \times B$ and $a \in A$, we denote the {\em image\/} of
$a$ under $R$ by $R(a) \df \{b \in B: a \mathrel{R} b\}$.  For $C
\subseteq A$ we define $R(C) \df \bigcup_{a \in C}R(a)$.  

The
\emph{inverse} $R^{-1}$ of a relation $R$, the identity relation
$\mathord{\id_A}$ on $A$, the relational composition $R \cdot S$ or
two relations $R,S$, the transitive closure $R^+$ of $R$, and the
reflexive-transitive closure $R^*$ of $R$ are defined in the usual
manner.

\paragraph*{Lposets.} 
For the rest of this paper let us fix a finite alphabet
$\Sigma$.  A {\em labeled partial order\/} (lposet, for short) over
$\Sigma$ is a structure $x = \<E, \mathord<, \lambda>$ where $E$ is a
finite set of {\em events}, $\mathord< \subseteq E \times E$ is an
(irreflexive) partial order, and $\lambda: E \rightarrow \Sigma$ is a
labeling function.

Let $x$ be a lposet and let $e_1, e_2 \in E_x$. We use the following
notions: The {\em reflexive closure\/} of $<_x$ is $\mathord{\leq_x} \df
\mathord<_x \cup \id_{E_x}$.  Unrelated events are called {\em
concurrent\/}, i.\,e., $e_1 \co_x e_2 \dfeq e_1 \not\leq_x e_2 \And e_2
\not\leq_x e_1$, while related events are {\em in line}: $e_1 \li_x
e_2 \dfeq e_1 <_x e_2 \Or e_2 <_x e_1$.

  The {\em downward closure\/} of a set
$D \subseteq E_x$ is $\dncl{D} \df \mathord{\leq^{-1}_x}(D)$. If $D =
\dncl{D}$ holds, then $D$ is called {\em downward closed\/} in $x$. By
$\cl(x)$ we denote the set of downward closed sets in $x$.  If $D
\subseteq E_x$, then $x[D] \df \<D, \mathord<_x \cap (D \times D),
\lambda\res D>$ is the lposet {\em generated by\/} $D$ in $x$ ($\lambda\res D$ denotes the
restriction of $\lambda$ to $D$).

\paragraph*{Pomsets.}
Lposets $x$ and $y$ over $\Sigma$ are called {\em isomorphic}, written $x \approx y$, if
there is a bijection $f: E_x \rightarrow E_y$ such 
that $(e_1 <_x e_2 \Leftrightarrow f(e_1) <_y f(e_2)) \And \lambda_x = \lambda_y \circ f$ holds.
A {\em partially ordered multiset\/} (a {\em pomset} for short) over $\Sigma$
is an isomorphism class of lposets, i.\,e., a set $[x] \df \{y: x
\approx y\}$.  We fix the convention, that pomsets are denoted by
boldfaced small letters $\mathbit{x}, \mathbit{y}, \mathbit{z}$.
Moreover $\mathbit{x}$ is assumed to be the equivalence class $[x]$ of
$x$. By this convention, $E_x$ always denotes the set of events of a
representative $x$ of $\mathbit{x}$.  The class of pomsets over
$\Sigma$ is denoted by $\Pomset(\Sigma)$.  

\begin{figure}[t]
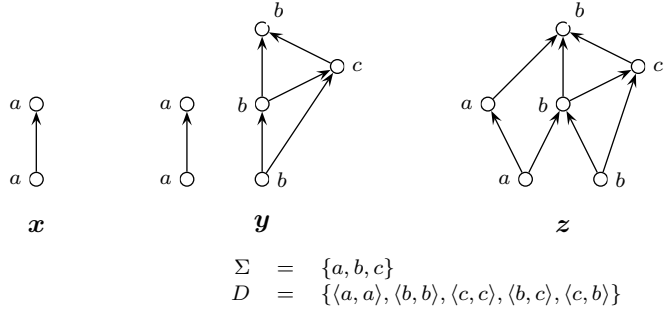

\begin{center}
\grNogrid 
\begin{gr}(0,-1)(8,3)

\grWindow[t](0,-.3){
\grEvent{a1}(0,1) \grEvLab[l]{$a$}
\grEvent{a2}(0,2) \grEvLab[l]{$a$}
\grLine{a1}{a2}
\grLine{c}{b3}
\grTextbox[b](0,.3){$\mathbit{x}$}
}

\grWindow[t](2,-.3){
\grEvent{a1}(0,1) \grEvLab[l]{$a$}
\grEvent{a2}(0,2) \grEvLab[l]{$a$}
\grEvent{b1}(1,1) \grEvLab[r]{$b$}
\grEvent{b2}(1,2) \grEvLab[l]{$b$}
\grEvent{b3}(1,3) \grEvLab[ur]{$b$}
\grEvent{c}(2,2.5) \grEvLab[r]{$c$}
\grLine{a1}{a2}
\grLine{b1}{b2}
\grLine{b1}{c}
\grLine{b2}{b3}
\grLine{b2}{c}
\grLine{c}{b3}
\grTextbox[b](1,.3){$\mathbit{y}$}
}

\grWindow[t](6,-.3){
\grEvent{a1}(.5,1) \grEvLab[l]{$a$}
\grEvent{a2}(0,2) \grEvLab[l]{$a$}
\grEvent{b1}(1.5,1) \grEvLab[r]{$b$}
\grEvent{b2}(1,2) \grEvLab[l]{$b$}
\grEvent{b3}(1,3) \grEvLab[ur]{$b$}
\grEvent{c}(2,2.5) \grEvLab[r]{$c$}
\grLine{a1}{a2}
\grLine{a1}{b2}
\grLine{b1}{b2}
\grLine{b1}{c}
\grLine{a2}{b3}
\grLine{b2}{b3}
\grLine{b2}{c}
\grLine{c}{b3}
\grTextbox[b](1,.3){$\mathbit{z}$}
}
\grTextbox[br](8,-1){\footnotesize$\begin{array}{rcl}
\Sigma &=& \{a, b, c\}\\
D &=& \{\<a, a>, \<b, b>, \<c, c>, \<b, c>, \<c, b>\}
\end{array}$} 
\end{gr}
\end{center}
\vspace{-.5cm}
\caption{\label{fig:pomsets}Example pomsets} 
\vspace{-.2cm}
\end{figure}

Fig. \ref{fig:pomsets} shows examples of pomsets. Graphically, we represent pomsets as directed acyclic
(not necessarily connected) graphs. Nodes are labeled with elements from the underlying alphabet $\Sigma$. 
Transitive arcs are sometimes omitted.

Let $\mathbit{x}, \mathbit{y} \in \Pomset(\Sigma)$ be pomsets. Then
$\mathbit{x}$ is called a {\em prefix\/} of $\mathbit{y}$---denoted
$\mathbit{x} \approximates \mathbit{y}$---iff there are
representatives $x \in \mathbit{x}$ and $y \in \mathbit{y}$ such that
$E_x \subseteq E_y \And \cl(x) \subseteq \cl(y)$ holds.  If there are
representatives $x \in \mathbit{x}$ and $y \in \mathbit{y}$ such that
$E_x = E_y \And \mathord<_x \subseteq \mathord<_y$ holds, then
$\mathbit{x}$ is called {\em less sequential\/} than $\mathbit{y}$.
This is denoted by $\mathbit{x} \seq \mathbit{y}$.  It is easy to see
that both $\approximates$ and $\seq$ partially order
$\Pomset(\Sigma)$. In Fig. \ref{fig:pomsets}, $\mathbit{x}
\approximates \mathbit{y}$, $\mathbit{x} \approximates \mathbit{z}$,
and $\mathbit{y} \seq \mathbit{z}$, holds.

An alternative definition of the prefix relation $\approximates$ and the ordering by
the degree of sequentiality $\seq$ can be obtained by introducing
the notion of {\em weak homomorphisms\/} between representatives of
pomsets \cite{deussen98a}.

Special pomsets that will be encountered in this paper are:
\begin{enumerate}
\item Letters $a = [\{a\}, \emptyset, a \mapsto a]$ for $a \in \Sigma$ (we abuse
      $a, b, c, \ldots$ to denote both letters from $\Pomset(\Sigma)$ and
      from $\Sigma$).

\item Strings $[\{0, \ldots, n-1\}, \mathord<, i \mapsto a_i]$ for $a_0a_1\ldots a_{n-1} \in \Sigma^*$, where $<$ denotes the standard order relation on integers.

\item The empty word $\epsilon = [\emptyset, \emptyset, \emptyset]$. 
\end{enumerate} 
In this paper we do not distinguish between strings and pomset strings, i.\,e, if
$\Sigma$ is an alphabet then $\Sigma^*$ is considered to be the set of pomsets $\sigma$ over
$\Sigma$ such that $<_\sigma$ is a total ordering.

If $\mathbit{x} \in \Pomset(\Sigma)$, then by $\lin(\mathbit{x}) \df \{\sigma \in \Sigma^*: \mathbit{x} \seq \sigma\}$
we denote the set of {\em linearizations\/} of $\mathbit{x}$. 
 
\paragraph*{Dependencies and Weak Sequential Composition.}
A reflexive and symmetric relation $D \subseteq \Sigma \times \Sigma$
is called a {\em dependence\/} on $\Sigma$; for the rest of this paper
let $D$ be a dependence on $\Sigma$.  If $\mathbit{x}$ and
$\mathbit{y}$ are lposets over $\Sigma$, such that $E_{\mathbit x} \cap E_{\mathbit y} =
\emptyset$ holds, then the {\em weak sequential composition\/}
$\mathbit{x} \wsc{D} \mathbit{y}$ is defined by
\begin{eqnarray*}
\mathbit{x} \wsc{D} \mathbit{y} &\df& [E_{\mathbit x} \cup E_{\mathbit y}, (\mathord<_{\mathbit x} \cup \mathord<_{\mathbit y} \cup R)^+, \lambda_{\mathbit x} \cup \lambda_{\mathbit y}],
\end{eqnarray*}
where $R \subseteq E_{\mathbit x} \times E_{\mathbit y}$ is given by $e_1 \mathrel{R} e_2
\dfeq \lambda_{\mathbit x}(e_1) \mathrel{D} \lambda_{\mathbit y}(e_2)$.

A pomset $\mathbit{x}$ is called {\em $D$-consistent\/} if we have,
for all $e_1, e_2 \in E_x$, $e_1 \co_x e_2 \Rightarrow
\neg\lambda_x(e_1) \mathrel{D} \lambda_x(e_2)$. Let $\Pomset(\Sigma,
D)$ denote the class of $D$-consistent pomsets. Clearly, if
$\mathbit{x}$ and $\mathbit{y}$ are $D$-consistent, then also
$\mathbit{x} \wsc{D} \mathbit{y}$ is.

Another operation on pomsets which is closely related to $\wsc{D}$ is
the {\em unsequentialization\/} via $D$: If $\mathbit{x} \in
\Pomset(\Sigma)$, then by $\unseq{\mathbit{x}}$ we denote the pomset
$[E_x, R^+, \lambda_x]$, where $R \subseteq E_x \times E_x$ is defined
by $e_1 \mathrel{R} e_2 \dfeq e_1 \leq e_2 \And \lambda_x(e_1)
\mathrel{D} \lambda_x(e_2)$.
\footnote{The operations $\wsc{D}$ and
  $\unseq{\cdot}$ impose an interesting and fruitful connection to the
  theory of Mazurkiewicz traces \cite{maz95}.  Although it is far
  beyond the scope of this paper it should be noted that pomsets of
  the form $\unseq{\mathbit{x}} \in \Pomset(\Sigma, D)$ are just
  alternative representations of Mazurkiewicz traces: in fact we have
  that $\lin(\unseq{\mathbit{x}})$ is a Mazurkiewicz trace over
  $\Sigma$ and $D$; moreover, the operation $\wsc{D}$ coincides with
  trace concatenation.  }

The pomsets in Fig. \ref{fig:pomsets} are all $D$-consistent for the dependence
$D$ shown in that figure. We have $\unseq{\mathbit{z}} = \mathbit{y}$.

The following lemma justifies the relation between the operations
$\wsc{D}$ and $\unseq{\cdot}$.

\begin{lemma}\label{lemma:unseq}
Let $\mathbit{x}, \mathbit{y}$ in $\Pomset(\Sigma)$.
Then $\unseq{\mathbit{x} \wsc{D} \mathbit{y}} = \unseq{\mathbit{x}} \wsc{D} \unseq{\mathbit{y}}$.
\end{lemma}

Some more definitions: if $A \subseteq \Sigma$ is a set of symbols and
$\mathbit{x}$ is a pomset, then $\mathbit{x}\res A \df
[x[\lambda^{-1}_x(A)]]$.  $\mathbit{x}\res A$ is called the {\em
  restriction of $\mathbit{x}$ to $A$}, i.e., $\mathbit{x}$ restricted
to those events labeled with elements from $A$. Finally, a set of
pomsets $X \subseteq \Pomset(\Sigma)$ is called {\em pre-closed\/} if
$\mathbit{x} \in X \And \mathbit{y} \approximates \mathbit{x}
\Rightarrow \mathbit{y} \in X$ holds.

\subsection{Partial Order Semantics for MSCs}

\setlength{\levelheight}{.5cm} \setlength{\inlineoverlap}{1.2cm}
\setlength{\conditionheight}{.4cm} \setlength{\instheadheight}{.4cm}
\setlength{\lastlevelheight}{.1cm}

\begin{figure}[t]
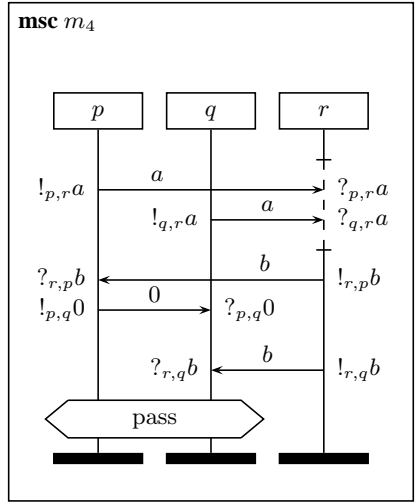

\begin{center}
\begin{footnotesize}
  \setmscvalues{small}

      \begin{msc}{\ensuremath{m_4}}
        \declinst{p}{}{$p$}
        \declinst{q}{}{$q$}
        \declinst{r}{}{$r$}
        \coregionstart{r} \nextlevel
        \mess{$a$~~~~~~~~~~~~~~~~~~}{p}{r} \startlabel[r]{$\send{p, r}a$}\endlabel[l]{$\rec{p, r}a$} \nextlevel
        \mess{$a$}{q}{r} \startlabel[r]{$\send{q, r}a$}\endlabel[l]{$\rec{q, r}a$} \nextlevel
        \coregionend{r} \nextlevel
        \mess{~~~~~~~~~~~~~~~~~~$b$}{r}{p} \startlabel[l]{$\send{r,p}
          b$} \endlabel[r]{$\rec{r,p} b$} \nextlevel
        \mess{$0$}{p}{q} \startlabel[r]{$\send{p, q}0$}\endlabel[l]{$\rec{p, q}0$} \nextlevel[2]
        \mess{$b$}{r}{q} \startlabel[l]{$\send{r, q}b$}\endlabel[r]{$\rec{r, q}b$} \nextlevel
        \condition{$\Vpass$}{p,q} \nextlevel
      \end{msc}

\end{footnotesize}
\end{center} 
\vspace*{-.5cm}
\caption{\label{fig:msc}An annotated MSC.}
\vspace*{-.2cm}
\end{figure}

To define the semantics of MSCs based on pomsets, we first need to fix
an alphabet $\Com\Sigma$ and a dependence $\Com D$ on $\Com\Sigma$.

\paragraph*{Communication Alphabet and Dependence.} 

Let $M$ be a set of {\em messages \/} and $P$ a set of {\em
  instances\/} fixed throughout this paper.  We assume that there is a
non-empty set $T \subset P$ of {\em port instances}; the instances
in $P - T$ will be called {\em SUT instances}.  Usually we will have
$|P - T| = 1$, but the theory presented in the following does not rely
on this.

Let $\Sigma_{\SendSym}, \Sigma_{\RecSym}$ 
be the two alphabets:
\begin{enumerate}
\item $\Sigma_{\SendSym} \df \{\SendSym\} \times P \times M \times P$
  is the set of {\em send actions}. Its elements $\<\SendSym, p, m,
  q>$ will be denoted by $\send{p, q}m$,
\item $\Sigma_{\RecSym} \df \{\RecSym\} \times P \times M \times P$ is
  the set of {\em receive actions}.  Its elements $\<\RecSym, p, m, q> \in
  \Sigma_?$ will be denoted by $\rec{p, q}m$.
\end{enumerate}

We put $\Com\Sigma \df \Sigma_{\SendSym} \cup \Sigma_{\RecSym}$ 
to be the set of {\em communications}. The mapping $\ins(a)$
identifies the {\em instance\/} of an action $a \in \Com\Sigma$,
i.\,e.,  $\ins(\send{p, q}m) \df p$ and $\ins(\rec{p, q}m) \df q$. We
put $\ObsAlph \df \{a \in \Com\Sigma: \ins(a) \in T\}$ to be the set
of {\em tester observable actions}. For convenience, we furthermore
define $\ObsSend \df \ObsAlph \cap \Sigma_{\SendSym}$ and $\ObsRec \df
\ObsAlph \cap \Sigma_{\RecSym}$.

Fig. \ref{fig:msc} gives a few examples of this syntax of actions. It
shows the expansion of the first alternative of $m_4$ where the
generalized ordering has been replaced by sending the void message
$0$. The messages have been annotated with the corresponding symbols
from $\Com\Sigma$.

To build pomsets from actions, we define the dependence $\Com D$ on
$\Com\Sigma$: let $\Com D \subseteq \Com\Sigma \times \Com\Sigma$ be
the smallest reflexive, symmetric relation containing:
\begin{itemize}
\item $\<a, b>$ with  $\ins(a) = \ins(b)$ and $a$ and $b$ are not
  placed on the same co-region,
\item $\<\send{p, q}m, \rec{p, q}m>$ for instances $p, q \in P$ and
  messages $m \in M$.
\end{itemize}

To keep things simple, we restrict ourself to the following MSC
operators: message sending and receiving, co-regions, and the
alternative inline expression, which allows the expression of optional
behavior and finite iterations.  We simulate general ordering by
sending a void message $0$, which might also be sent between two
port or SUT instances.  Conditions are only allowed to assign
verdicts and are not dealt with by the semantics. In order to obtain a
set of $\Com D$-consistent pomsets, we have to impose the restriction
that identical actions (e.\,g. sending of a message twice from an
instance $p$ to an instance $q$) are not placed on the same co-region.

The semantics of an MSC $\mathscr{M}$ is given by a pre-closed set
of pomsets $X_{\mathscr{M}} \subseteq \Pomset(\Com\Sigma, \Com D)$.
We illustrate the construction of $X_{\mathscr{M}}$ only by informal
means of an example ($m_3$ from Fig.~\ref{fig:alternatives}); the
translation is done similar to \cite{kat-lam98} with slightly
different syntax for events of pomsets.

In the following, $\wsc\ComSym$ abbreviates $\wsc{\Com{D}}$.


The semantics of our example $m_3$  is  given by the set
$X_{m_3}$:
\begin{eqnarray*}
X_{m_3} &=& \{\mathbit{z} \in \Pomset(\Com\Sigma, \Com D): \mathbit{z} \approximates \mathbit{x} \wsc{\ComSym} \mathbit{y}_{1} \Or \mathbit{z} \approximates \mathbit{x} \wsc{\ComSym} \mathbit{y}_{2}\}.
\end{eqnarray*}
where $\mathbit{x}, \mathbit{y}_1$, and $\mathbit{y}_2$ are defined by:
\begin{eqnarray}
\mathbit{x} &\df& \send{p,r}a \wsc{\ComSym} \send{q,r}a \wsc{\ComSym} \rec{p,r}a \wsc{\ComSym} \rec{q,r}a \label{eqn:x}\\
\mathbit{y}_1 &\df& \send{r,p}b \wsc{\ComSym} \rec{r,p}b \wsc{\ComSym} \send{p,q}0 \wsc{\ComSym} \rec{p,q}0 \wsc{\ComSym} \send{r,q}b \wsc{\ComSym} \rec{r,q}b \label{eqn:y1}\\
\mathbit{y}_2 &\df& \send{r,q}b \wsc{\ComSym} \rec{r,q}b \wsc{\ComSym} \send{q,p}0 \wsc{\ComSym} \rec{q,p}0 \wsc{\ComSym} \send{r,p}b \wsc{\ComSym} \rec{r,p}b \label{eqn:y2}
\end{eqnarray}

Without a proof (which would require a more formal treatment of the
definition of $X_{\mathscr{M}}$ we state:

\begin{lemma}\label{lemma:msc-pomsets-are-least-sequential}
If $\mathscr{M}$ is a MSC, then $\xunseq{\mathbit{x}}{D_\ComSym} = \mathbit{x}$ for all $\mathbit{x} \in X_{\mathscr{M}}$.
\end{lemma}

\subsection{Message Sequence Charts as Test Purposes}

Now that we have explained how to assign semantics to an MSC, we
show how MSCs can be utilized as a formal language to express test
purposes. We discuss how the notion of a test verdict can be
integrated into an MSC and how it can be guaranteed that an MSC
specifies behavior that is amenable to black-box testing.

\paragraph*{Verdict assignments.}

Syntactically, a verdict assignment is expressed by a condition on the
port instances on the very end of each terminal alternative of the
MSC. Semantically, the condition-like constructs pass, fail, and
inconc are not treated as an ordinary condition but as a convenient
way to define a \emph{verdict assignment}: \footnote{Alternatively,
  one could allow verdict conditions to appear also at other places
  within the MSC and, e.g., use the verdict assignment rules of TTCN-3
  \cite{etsi01:TTCN3} to resolve the case where different verdicts are
  encountered during a single run through the MSC.}

Let  $\Verdicts \df \{\Vpass, \Vfail, \Vinconc, \Vnone\}$ be a set of
{\em verdicts\/} and let $\FinVerdicts \df V -\{\Vnone\}$ be the set
of {\em final verdicts}.  A mapping $\ver: X \rightarrow V$ for some
finite, pre-closed set of pomsets $X$ is called a {\em verdict
  assignment\/} to $X$ if, for all $\mathbit{x} \in X$, we have:
\begin{enumerate}
\item $\exists \mathbit{y} \in X. \mathbit{x} \approximates
  \mathbit{y} \And \ver(\mathbit{y}) \neq \Vnone$, i.\,e., every pomset
  can be extended to a pomset that is assigned a final verdict, and
\item $\ver(\mathbit{x}) \neq \Vnone \Rightarrow \forall \mathbit{y}
  \in X. \mathbit{x} \not< \mathbit{y}$, i.\,e., pomsets that are
  assigned a final verdict are maximal in $X$.

\end{enumerate}

The verdict conditions drawn in an MSC $\mathscr{M}$ are used to define
a verdict mapping $\ver_{\mathscr{M}}$. Again, we introduce this
informally by the example of $m_3$ from Fig.~\ref{fig:alternatives},
where $\ver_{m_1}$ is defined by:
\begin{eqnarray*}
\ver_{m_3}(\mathbit{z}) &=& \begin{cases}
                           \Vpass, & \text{if } \mathbit{z} = \mathbit{x} \wsc{\ComSym} \mathbit{y}_{1};\\
                           \Vfail, & \text{if } \mathbit{z} = \mathbit{x} \wsc{\ComSym} \mathbit{y}_{2};\\
                           \Vnone, & \text{otherwise.}
                   \end{cases}
\end{eqnarray*}

It is obvious that not every MSC that satisfies the syntactic
restrictions that have been introduced above constitutes a test
purpose, i.\,e., describes behavior of the SUT that can be tested in a
black box testing approach. For example consider a modification of
$m_3$ from Fig.~\ref{fig:alternatives}, where the generalized ordering
constraints have been eliminated. There the verdict does not depend on
the order in which the messages $b$ can be observed at the ports of
the SUT but rather on the (SUT-internal) events that cause these
messages to be sent. Clearly, such an event is not visible to a
black-box test system and hence no test case can distinguish between
the behavior of the first and second alternative. In the following we
present a number of criteria that an MSC must satisfy to be
considered a \emph{well-formed} test purpose. Later we will see that
these criteria indeed guarantee the existence of a valid test case for
a test purpose.

\paragraph*{Well-Formed Test Purposes.}
First, we define a function that reduces the semantics of an MSC to
the information that is available to the test case, i.\,e., the sequences of
events that occur on port instances:

Given test purpose MSC $\mathscr{M}$ with semantics $X_{\mathscr{M}}$.
For $\mathbit{x} \in X_{\mathscr{M}}$, we define the \emph{tester
  observable traces} of $\mathbit{x}$ by $\obs(\mathbit{x}) \df
\lin(\mathbit{x} \res \ObsAlph)$.

A MSC $\mathscr{M}$ is
called a \emph{well-formed test purpose} if it is possible to determine its state (and
hence assigned verdict) based on this information in its tester observable traces, i.\,e., if
\begin{enumerate}\def\theenumi{$\mathrm{WF}_{\text{\arabic{enumi}}}$}

\item\label{enum:wf2} for every $\mathbit{x}, \mathbit{y} \in X_{\mathscr{M}} \res
  \ObsAlph$, $\lin(\mathbit{x}) \cap \lin(\mathbit{y}) \neq \emptyset$ implies
  $\mathbit{x} = \mathbit{y}$.
\end{enumerate}

Unfortunately, this restriction does not yet suffice to guarantee that
an MSC describes testable behavior. Another aspect that needs
considerations is which party resolves \emph{essential choice} in the
sense of the following definition:

Let $X \subseteq \Pomset(\Com\Sigma, \Com D)$ be a pre-closed set of
pomsets. A pomset $\mathbit{x} \in X$ is called a \emph{choice point}
for two actions $a, b \in \Com\Sigma$ in $X$ if $\mathbit{x}
\wsc{\ComSym} a \in X$, $\mathbit{x} \wsc{\ComSym} b \in X$, and

\begin{eqnarray*}
\left\{\mathbit{y} \in \max{}_\approximates(X): \mathbit{x} \wsc{\ComSym} a \leq \mathbit{y}\right\}
   &\neq& \left\{\mathbit{y} \in \max{}_\approximates(X): \mathbit{x} \wsc{\ComSym} b \leq \mathbit{y}\right\} ,
\end{eqnarray*}
where $ \max_\approximates(X)$ denotes the $\approximates$-maximal pomsets in $X$.

Coming back to example from Fig.~\ref{fig:alternatives} with semantics
$X_{m_3}$ as defined in (\ref{eqn:x})--~(\ref{eqn:y2}), $\mathbit{x}
\res \ObsAlph$ is a choice point for $\rec{r,p}b$ and $\rec{r,q}b$. On
the other hand, $\epsilon$ is not a choice point even though there are
two ``available'' communications, namely $\send{p,r}a$ and
$\send{q,r}a$, since this choice does not alter the reachable maximal
configurations.

We require, for a well-formed test purpose, that each  choice point
is resolved by a message from the SUT:
\begin{enumerate}\setcounter{enumi}{1}\def\theenumi{$\mathrm{WF}_{\text{\arabic{enumi}}}$}
\item\label{enum:wf3} If $\mathbit{x}$ is a choice point of
  $X_{\mathscr{M}} \res \ObsAlph$ for actions $a, b \in \ObsAlph$,
  then both $a, b \in \ObsRec$.
\end{enumerate}

This restriction is necessary because both other possibilities for a
choice point ($a,b \in \ObsSend$ or $a \in \ObsSend$ and $b \in
\ObsRec$) are undesirable in a test purpose:
a choice that has to be resolved by the test case indicates
  that the test purpose should indeed be (at least) two test purposes,
  one for each choice of the test case. Otherwise, a deterministic
  test case will only be able to test the part of the test purpose
  that corresponds to the (necessarily fixed) way the test case
  resolves the choice.
On the other hand, a choice that can be resolved simultaneously by the test case and
  SUT leads to problems because it might lead to a race condition
  where both test case and SUT resolve the choice in an inconsistent
  manner. This situation bears strong resemblance to the presence of
  non-local choice in the MSC~\cite{benabdallah97syntactic}.

\begin{figure}[t]
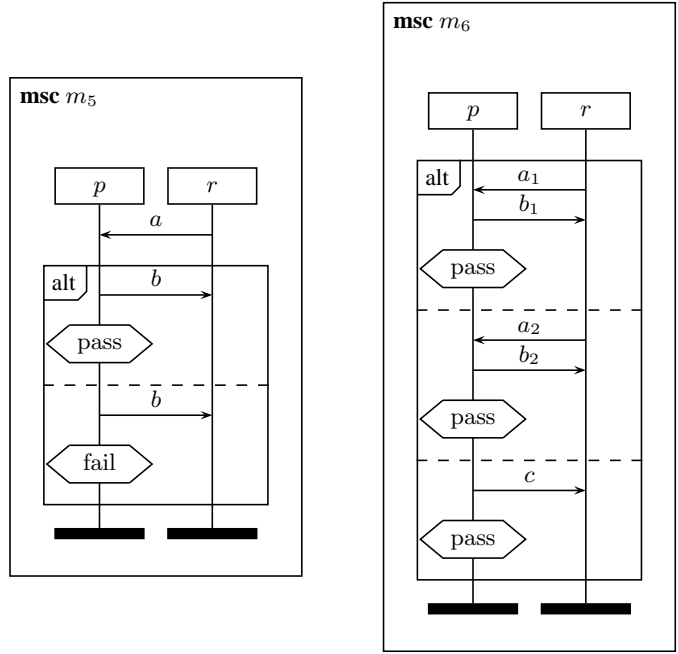

  \centering
  \begin{footnotesize}
    \setmscvalues{small}
    \begin{tabular}{m{.34\linewidth}cm{.35\linewidth}}
      \begin{msc}{\ensuremath{m_5}}
        \declinst{p}{}{$p$}
        \declinst{r}{}{$r$}
        \mess{$a$}{r}{p} \nextlevel
        \inlinestart{alt}{alt}{p}{r} \nextlevel
        \mess{$b$}{p}{r} \nextlevel
        \condition{$\Vpass$}{p} \nextlevel[2]
        \inlineseparator{alt} \nextlevel
        \mess{$b$}{p}{r} \nextlevel
        \condition{$\Vfail$}{p} \nextlevel[2]
        \inlineend{alt}
      \end{msc}
      & &
      \begin{msc}{\ensuremath{m_6}}
        \declinst{p}{}{$p$}
        \declinst{r}{}{$r$}
        \inlinestart{alt}{alt}{p}{r} \nextlevel
        \mess{$a_1$}{r}{p} \nextlevel
        \mess{$b_1$}{p}{r} \nextlevel
        \condition{$\Vpass$}{p} \nextlevel[2]
        \inlineseparator{alt} \nextlevel
        \mess{$a_2$}{r}{p} \nextlevel
        \mess{$b_2$}{p}{r} \nextlevel
        \condition{$\Vpass$}{p} \nextlevel[2]
        \inlineseparator{alt} \nextlevel
        \mess{$c$}{p}{r} \nextlevel
        \condition{$\Vpass$}{p} \nextlevel[2]
        \inlineend{alt}
      \end{msc}
    \end{tabular}
  \end{footnotesize}
  \vspace*{-.2cm}
  \caption{Two malformed MSCs}
  \vspace*{-.2cm}
  \label{fig:malformed-mscs}
\end{figure}
  
Figure~\ref{fig:malformed-mscs} shows examples of malformed MSCs: in
$m_5$ exist $\mathbit{x}, \mathbit{y} \in X_{m_5}
\res \ObsAlph$ with $\mathbit{x} \neq \mathbit{y}$ and $\sigma =
\send{r,p}a \cdot \rec{p,r}b \in\lin(\mathbit{x}) \cap \lin(\mathbit{y})
\neq \emptyset$, and hence $\mathrm{WF}_1$ is violated. Indeed there
exist $\mathbit{x},\mathbit{y}$ with that property such that
$\ver_{m_5}(\mathbit{x}) = \Vpass$ and
$\ver_{m_5}(\mathbit{y}) = \Vfail$. Taking into account the
fact that a test system will only observe $\sigma$ it is clear that
$m_5$ does not describe testable behaviour---which verdict
should a test system assign after observing $\sigma$?  The MSC
$m_6$ is malformed because it violates $\mathrm{WF}_2$:
$\epsilon$ is a choice point for the actions $\send{r,p} a_1$,
$\send{r,p} a_2$, and $\send{p,r} c$. In its initial configuration,
the test system can either (deterministically) send $a_1$ or $a_2$,
but will then not be able to test the behaviour of the SUT that
corresponds to the respective other choice. Also, what happens if the
test system decides to perform action $\send{r,p} a_1$ while the SUT,
before it has received $a_1$, performs $\send{p,r} c$?  This
behaviour is not defined by the MSC.  For an example of a well-formed
MSC, the reader may verify that $m_3$ from Fig.~\ref{fig:alternatives}
is indeed well-formed.

\section{Test Case  Validity}
\label{section:validity}

We now define the validity of a test case w.\,r.\,t.\ a well-formed test
purpose. Our definition is different from the available conformance relations
for labeled transition systems because it assigns different roles to
test case and SUT.  We show that the well-formedness conditions on
MSCs from the previous section suffice to guarantee the existence of a
valid test case.  Moreover, we give a simple decision procedure that
decides validity of a test case and prove its correctness.

First, we need to formalize the notion of a test case. Intuitively,
a test case interacts with the SUT by means of exchanging messages
and finally assigning a verdict.  Formally, we model a test case as
follows:

\paragraph*{Test Cases.} A \emph{test case} is a partial function $\TS : \ObsAlph^* \pfun
\ObsSend \mathrel{\dot \cup} \{ \delta \} \mathrel{\dot \cup}
\FinVerdicts$, where $\delta$ is a symbol that denotes
\emph{quiescence} of the test case.

A \emph{run} of a test case $\TS$ is a sequence $\sigma_0, \sigma_1, \ldots, \sigma_n$,
of words from $\ObsAlph^*$ such that $\sigma_0 = \epsilon$, 
and $\sigma_i \xrightarrow[\TS]{} \sigma_{i + 1}$ for $0 \leq i < n$, where the
relation $\xrightarrow[\TS]{}$ is defined by
\begin{gather*}
\sigma \xrightarrow[\TS]{} \sigma a \dfeq 
\TS(\sigma)\ \text{defined} \And \left(\TS(\sigma) = a \in \ObsSend \Or
  \TS(\sigma) = \delta \And a \in \ObsRec\right)
\end{gather*}
A run $\sigma_0, \sigma_1, \ldots, \sigma_n$ is called
\emph{complete} if $\TS(\sigma_n) \in \FinVerdicts$.
Note that it is indeed impossible to extend a complete run due to the
definition of $\xrightarrow[\TS]{}$.

In the following we will show how to model test case validity as a
certain language inclusion problem.

\paragraph*{Test Languages.} 
Both the runs of a test case and the tester observable traces of a
well-formed test purpose naturally induce \emph{test languages}, i.\,e.,
languages $L \subseteq \ObsAlph^*$ together with a verdict assignments $\ver_L$:

For a test case $\TS$, the test language $\<L_\TS, \ver_\TS>$ is
defined by $L_\TS \df \{ \sigma \in \ObsAlph^* : \epsilon
\xrightarrow[\TS]{}^* \sigma \}$ with verdict assignment $\ver_\TS$
defined by
$$\ver_\TS(\sigma) \df \begin{cases} \TS(\sigma), & \text{if}\ 
  \TS(\sigma)
  \in \FinVerdicts; \\
  \Vnone, & \text{otherwise}.
\end{cases}
$$

For a well-formed test purpose $\mathscr{M}$, the induced test
language $\<L_{\mathscr{M}}, \ver_{\mathscr{M}}>$ is defined by
setting $L_{\mathscr{M}} \df \obs(X_{\mathscr{M}})$ and, for $\sigma
\in L_{\mathscr{M}}$, $\ver_{\mathscr{M}}(\sigma) \df
\ver_{\mathscr{M}}(\mathbit{x})$ for the (due to \ref{enum:wf2} uniquely
defined) $\mathbit{x} \in X_{\mathscr{M}}$ with $\sigma \in
\obs(\mathbit{x})$.

It can easily be shown that $v_\TS$ and $v_{\mathscr{M}}$ are
well-defined and satisfy the requirements imposed on verdict
assignments.

What is the correct relation between $\<L_{\mathscr{M}},
\ver_{\mathscr{M}}>$ and $\<L_{\TS}, \ver_{\TS}>$ to define validity
of $\TS$ w.\,r.\,t.\ $\mathscr{M}$? Clearly, $\ver_{\mathscr{M}}$ and
$\ver_{\TS}$ should agree on $L_{\mathscr{M}} \cap L_{\TS}$. But what
is the right relations between $L_{\mathscr{M}}$ and $L_{\TS}$? None
of the ``obvious'' choices leads to a satisfactory notion of validity:
\begin{itemize}
\item if we would require $L_{\mathscr{M}} \subseteq L_{\TS}$ then
  there would be no valid test cases for any test purpose that allows
  (inessential) choice between two actions $a,b \in \ObsSend$ because
  $L_{\mathscr{M}}$ contains traces for both  choices while a
  deterministic test case would be limited to only a single choice.
\item requiring $L_{\TS} \subseteq L_{\mathscr{M}}$ would allow the
  test case to send arbitrary messages to the SUT even though these
  would not be specified in the test purpose
\item if we require $L_{\TS} \cap L_{\mathscr{M}} \neq \emptyset$ then
  the test case would only be required to react to one of the possible
  many (essential) choices that the SUT might have.
\end{itemize}

While the first option matches the intuitive meaning of test case
validity best, it needs to be modified to eliminate the influence of
inessential choice. This is done by means of the following equivalence
relation on strings:

Let $L \subseteq \ObsAlph^*$ be a language.  We define an equivalence
relation $\mathord\simeq_{L} \subseteq L \times L$ by setting
$\sigma \simeq_{L} \rho \dfeq \text{$\rho$ is a permutation of $\sigma$ such that 
$\sigma \res \ObsRec = \rho \res \ObsRec$}$.
The {\em equivalence class\/} w.\,r.\,t. $\simeq_{L}$ of $\sigma \in
L$ is denoted by $[\sigma]_{L} \df \{\rho \in L: \sigma \simeq_{L}
\rho\}$.

\paragraph*{Test Case Validity.} 
Let $\mathscr{M}$ be a well-formed  test purpose
and $\TS$ be a test case for $\mathscr{M}$. Then $\TS$ is called a
\emph{valid} test case w.\,r.\,t.\ $\mathscr{M}$ if
\begin{itemize}
\item for every $\sigma \in L_{\TS} \cap L_{\mathscr{M}}$,
  $\ver_{\TS}(v) = \ver_{\mathscr{M}}(v)$, and
\item for every $\sigma \in L_{\mathscr{M}}$ with
  $\ver_{\mathscr{M}}(\sigma) \in \FinVerdicts$,
  $[\sigma]_{L_{\mathscr{M}}} \cap L_{\TS} \neq \emptyset$.
\end{itemize}

Since we have given the definition both for well-formed test purposes
and test case validity, it would be futile to use one to justify the
other.  What can be shown formally though, is that these notions are
compatible in the following sense:

\begin{theorem}\label{theo:valid-tester-exists}
  Let $\mathscr{M}$ be a well-formed test purpose. Then there exists a
  test case $\TS$ that is valid w.\,r.\,t.\ $\mathscr{M}$. $\TS$ can be
  computed effectively from $\mathscr{M}$.
\end{theorem} 

Also, it is easy to see that there are MSCs that violate
one of the well-formedness conditions, for which no valid test case
exists.

\paragraph*{Deciding Validity.}

In the following we present an algorithm that decides validity of a
test case w.\,r.\,t.\ a well-formed test purpose and establish the
algorithm's correctness. Interestingly, the algorithm does not require
the calculation of the $\simeq_{L_{\MS}}$-classes but only refers to
$\obs(X_{\MS})$, $\ver_\MS$, and $L_\MS$, which can easily be derived
from $\MS$.

\medskip

\begin{algo}
\small
\begin{tabbing}
\quad\=\quad\=\quad\=\quad\=\quad\=\quad\=\quad\=\quad\=\quad\=\kill
\>{\bfseries valid}$(${\it test\_purpose}  $\mathscr{M}$; {\it test\_case} $\TS$; {\it string} $\rho)$ \{\\
\>\>    \textbf{if} $\TS(\rho)$ \textit{is undefined} \textbf{then fail};\\
\>\>    \textbf{if} $\TS(\rho) \in \FinVerdicts \And \TS(\rho) \neq \ver_{\mathscr{M}}(\rho)$ \textbf{then fail};\\
\>\>    \textbf{else if} $\TS(\rho) = \delta$ \textbf{then}\\
\>\>\>      \textbf{if} {\it en}$(\mathscr{M},\rho) \cap \ObsRec = \emptyset$ \textbf{then fail};\\
\>\>\>      \textbf{else foreach} $a \in$ {\it en}$(\mathscr{M},\rho) \cap \ObsRec$ \textbf{do} {\it valid}$(\MS, \TS, \rho \cdot a)$;\\
\>\>   \textbf{else if} $\rho \cdot \TS(\rho) \not\in L_{\mathscr{M}}$ \textbf{then fail};\\
\>\>    \textbf{else} {\it valid }$(\MS, \TS, \rho \cdot \TS(\rho))$;\\
\>\>    \textbf{success};\\
\>$\}$\\
\>where {\it en}$(\mathscr{M},\rho) \df \{ a \in \ObsAlph : \rho \cdot a \in \obs(X_{\mathscr{M}}) \}$
\end{tabbing}
\vspace*{-.5cm}
\caption{\label{algo:valid}Validation algorithm.}
\end{algo}

\begin{theorem}\label{theorem:alg-correct}
  Let $\mathscr{M}$ a well-formed test purpose and
  $\TS$ a test case for $\mathscr{M}$. Then $\TS$ is valid w.\,r.\,t.\
  $\mathscr{M}$ iff {\it valid}$(\mathscr{M}, \TS, \epsilon)$ does not
  fail.
\end{theorem}

\section{Practical Considerations}
\label{section:implementation}

The previous sections have discussed formally the relationship between
a test purpose defined using MSC and a test system that implements the
test purpose. No assumptions have been made on the test system besides
that it is deterministic and that it has observable test events and a
final verdict status.  An MSC based validator tool has been designed
and developed within a joint project between Nokia Research Center and
Fraunhofer FOKUS.

The validator is designed to run against any test system that provides
some basic functionality, like starting of a test case, retrieving the
status of the final verdict, sending and receiving messages, etc. The basic
idea was to create a validator that is not only able to validate the
abstract test suite but also a real test system (tester), i.\,e., an
abstract test suite plus its execution environment plus the glue that
is necessary to tie the test suite to the actual system under test.
Since this glue can be of considerable complexity, e.g., consisting of
implementations of various protocol stacks, message en- and decoders,
possibly tailored hardware, etc., testing of the whole test system is
indeed an important aspect.

This is also one of the advantages of our approach as compared to other
approaches like an isolated verification of the abstract test suite or
an automatic generation of test cases from test purposes.

Given a sufficiently detailed specification of the test purpose, a
combination of automatic generation of test cases
\cite{schmitt00:_test_gener_autol_testc} from the test purpose
together with a validation following our approach seems optimal. The
validation guarantees both the correctness of the implementation of
the generation algorithm and of the additional components that make up
the test system.

The design of the validator aims to make it as independent of the test
system as possible by defining a small, well-defined interface to
connect the validator to the test system. In our case study we have
used a TTCN-3 test system with the MSC validator. The validator accesses
the test system at its (proprietary) control interface to trigger the
execution of testcases and retrieve the final verdict. It uses
TTCN-3's standardized communication interface toward the
SUT~\cite{schhulz02:_implem_ttcn_test_system_tri} to exchange messages
with the test system. The MSC validator has been implemented using
JAVA and the test system runs independently of the validator.
Although not all work within this project has been completely
finished, results so far show that using MSCs as test purpose
definition language and as basis for the test case validation can
improve the quality of test cases and thus the quality of system
implementations.


\section{Future Work}
\label{section:outlook}

This paper defines a novel approach to test case validation and
provides the necessary theoretical background. Yet, it is only a first
step toward a working test case validation system. In particular, the
following issues need to be addressed in the future:

\paragraph*{Algorithms and Complexity.}

Deciding well-formedness of an MSC $\mathscr{M}$ so far requires the
calculation of the semantics $X_{\mathscr{M}}$, which is a costly
operation. A syntactic characterization of well-formedness would be
desirable because it would probably allow for faster tests for
well-formedness that could, e.g., also be built into an MSC editor to
support test purpose development by pointing out problematic
constructs. Additionally, a detailed analysis of the complexity of
well-formedness and test case validity would be desirable.

\paragraph*{Data.}

Since its last revision, data is an integral part of the MSC language.
An extension of our approach that also takes into account data passed
in messages is essential for the practical applicability of our
approach to a wider class of test cases. While this should not impose
any theoretical problems, it will be a challenge to integrate data
into the implementation in a user-friendly manner.

\paragraph*{Time.}

MSCs allow to express various timing constraints and timing aspects
are important in many testing efforts. Therefore, we plan to extend our
approach to MSCs with timing constraints. From a theoretical point of
view, this is probably the most interesting way to continue the work
presented in this paper.

\section{Appendix}

This appendix contains the proofs of
Theorems~\ref{theo:valid-tester-exists} and \ref{theorem:alg-correct}
(for technical reason in reverse order).  In the following, let
$\mathscr{M}$ denote a well-formed test purpose.

From property \ref{enum:wf2} we get that the function $\delin(\cdot):
\obs(X_{\mathscr{M}}) \rightarrow X_{\mathscr{M}}$
that maps
every $\sigma \in \obs(X_{\mathscr{M}})$ to a $\mathbit{x}_\sigma \in
X_{\mathscr{M}}$ such that $\sigma \in \lin(\mathbit{x}_\sigma)$ is in
fact a well-defined and total. It is easy to show the following
property:

\begin{lemma}\label{lemma:pre-invariance}
  Let $\mathscr{M}$ be a well-formed test purpose and $\rho,\sigma \in
  \obs(X_{\mathscr{M}})$ with $\rho \leq \sigma$. Then $\delin(\rho)
  \leq \delin(\sigma)$.
\end{lemma}

Let $\mathscr{M}$ be a well-formed test purpose, $\TS$ a test case for
$\mathscr{M}$ and $\sigma \in L_{\mathscr{M}}$ with
$\ver_{\mathscr{M}}(\sigma) \neq \Vnone$. A \emph{validation} for
$\sigma$ is a complete run $\rho_0 \xrightarrow[\TS]{} \rho_1
\xrightarrow[\TS]{} \cdots \xrightarrow[\TS]{} \rho_n$ such that
$\rho_n \simeq_{L_{\mathscr{M}}} \sigma$ and $\ver_\TS(\rho_n) =
\ver_{\mathscr{M}}(\sigma)$.
  
It is easy to see that validity of a test case w.\,r.\,t.\ a test purpose can
equivalently be formulated as follows.

\begin{lemma}\label{lemma:validation-run}
  Let $\mathscr{M}$ be a well-formed test purpose and $\TS$ a test
  case for $\mathscr{M}$. Then $\TS$ is valid w.\,r.\,t.
  $\mathscr{M}$ iff every $\sigma \in L_{\mathscr{M}}$ has a
  validation.
\end{lemma}

We will need the following technical lemma:

\begin{lemma}\label{lemma:prefer-send-action}
  Let $\mathscr{M}$ be a well-formed test purpose,
  $a, b \in \ObsAlph$ be actions, and $\rho,\sigma \in \ObsAlph^*$.
  Moreover, assume $\rho a, \rho b \in L_{\mathscr{M}}$,
  $\ver_{\mathscr{M}}(\sigma) \neq \Vnone$, and
  $\rho b \leq \sigma$. If $a \not\in \ObsSend$ or $b \not\in \ObsSend$ (or both),
  then there exists a $\sigma' \in
  L_{\mathscr{M}}$ with $\sigma \simeq_{L_{\mathscr{M}}}
  \sigma'$, $\ver_{\mathscr{M}}(\sigma) =
  \ver_{\mathscr{M}}(\sigma')$, $\rho a \leq \sigma'$, and
  $\ver_{\mathscr{M}}(\sigma) = \ver_{\mathscr{M}}(\sigma')$.
\end{lemma}

\begin{proof}
  Let $a,b,\rho,\sigma$ as required by the lemma and let $\mathbit{x}
  = \delin(\rho)$, and $\mathbit{z} = \delin(\sigma)$.  Let
  $\mathbit{y}_a, \mathbit{y}_b \in X_{\mathscr{M}} \res \ObsAlph$ such
  that $\mathbit{y}_a = \mathbit{x} \wsc{\ComSym} a$ and
  $\mathbit{y}_b = \mathbit{x} \wsc{\ComSym} b$. From \ref{enum:wf2} we
  get $\mathbit{x} \leq \mathbit{y}_b \leq \mathbit{z}$. From
  \ref{enum:wf3}, also $\mathbit{x} \leq \mathbit{y}_a \leq
  \mathbit{z}$ holds. Hence, there exists $\mathbit{u} \in
  X_{\mathscr{M}} \res \ObsAlph$ with $\mathbit{u} = \mathbit{x}
  \wsc{\ComSym} a \wsc{\ComSym} b$ and $\mathbit{u} \leq
  \mathbit{z}$ and we obtain $\sigma'$ setting $\sigma' = \rho a b
  \eta$, where $\eta$ is the string that can be appended to $\rho b$
  to obtain $\sigma$ with the first occurrence of $a$ deleted.  From
  what have said before, $\sigma'$ is a linearization of
  $\mathbit{z}$, hence $\sigma \simeq_{L_{\mathscr{M}}} \sigma'$ 
  and $\ver_{\mathscr{M}}(\sigma) =
  \ver_{\mathscr{M}}(\sigma')$.
\end{proof}

\begin{proof}[Proof of Theorem \ref{theorem:alg-correct}]
 Assume
  that {\bit valid}$(\mathscr{M},\TS,\epsilon)$ does not fail and let
  $\sigma \in L_{\mathscr{M}}$ with $\ver_{\mathscr{M}}(\sigma) \neq
  \Vnone$ and $|\sigma| = n$. By Lemma~\ref{lemma:validation-run}, it
  suffices to show that there exists a validation of $\sigma$. To this
  purpose, we will construct sequences $\rho_0,\dots, \rho_n$ and
  $\sigma_0, \dots, \sigma_n$ such that $|\rho_i| = i$, {\bit
    valid}$(\mathscr{M}.\TS,\rho_i)$ is called during the execution of
  the algorithm, $\rho_i \leq \sigma_i$, $\sigma_i
  \simeq_{L_\mathscr{M}} \sigma$, and
  $\ver_{\mathscr{M}}(\sigma) = \ver_{\mathscr{M}}(\sigma_i)$, for
  each $0 \leq i \leq n$.
  
  We start with $\rho_0 = \epsilon$ and $\sigma_i = \sigma$, which
  satisfies all the required properties. Assume that the sequences
  have been constructed up to $i$. Since {\bit
    valid}$(\mathscr{M},\TS,\rho_i)$ does not fail, $\rho_i \in
  L_{\mathscr{M}}$ holds and there are the following possibilities:
  \begin{itemize}
  \item $\TS(\rho) \in \FinVerdicts \And \TS(\rho) =
    \ver_{\mathscr{M}}(\rho)$. In this case, $|\rho| = n$ must hold
    because otherwise $\ver_{\mathscr{M}}(\sigma_i) \neq \Vnone$ and
    $\ver_{\mathscr{M}}(\rho_i) \neq \Vnone$, together with $\rho_i <
    \sigma_i$, would be (by Lemma~\ref{lemma:pre-invariance}) a
    contradiction to the fact that $\ver_{\mathscr{M}}$ is a verdict
    assignment on $L_{\mathscr{M}}$.
  \item $\TS(\rho) = \delta \And \ ${\bit en}$(\MS,\rho) \cap \ObsRec \neq
    \emptyset$. Then $i < n$ must hold and since $\rho_i < \sigma_i$,
    there exists $b \in \Ev(T)$ such that $\rho_i b \leq \sigma_i$. If
    $b \in \ObsRec$ then there will be a call {\bit
      valid}$(\mathscr{M},\TS,\rho_i b)$ and we set $\rho_{i+1} \df
    \rho_i b$ and $\sigma_{i+1} \df \sigma_i$ to continue the
    sequences.  Clearly, this satisfies all necessary properties. If
    $b \in \ObsSend$ then let $a \in \text{\bit en}(\MS,\rho_i) \cap
    \ObsRec$. Then $a, b, \rho_i, \sigma_i$ satisfy the
    prerequisites of Lemma~\ref{lemma:prefer-send-action}, which
    yields the existence of $\sigma_i' \in L_\mathscr{M}$ with
    $\sigma_i \simeq_{L_{\mathscr{M}}} \sigma_i'$,
    $\ver_{\mathscr{M}}(\sigma_i) =\ver_{\mathscr{M}}(\sigma_i')$, and
    $\rho_i a \leq \sigma_i'$. If we set $\rho_{i+1} \def \rho_i a$
    and $\sigma_{i+1} \df \sigma_i'$ then we have extended the
    sequence as required.
    \item $\TS(\rho) = a \in \ObsSend \And \rho a \in
      L_{\mathscr{M}}$. In this case we necessarily have to set
      $\rho_{i+1} \df \rho_i a$ and we need to show the existence of a
      suitable $\sigma_{i+1}$. This can be done similar to the
      previous case using Lemma~\ref{lemma:prefer-send-action}.
  \end{itemize}
  
  It is easy to see that $\TS(\rho_n) \in \FinVerdicts$ and by
  construction it holds that $\rho_n \simeq_{L_{\mathscr{M}}}
  \sigma$ as well as $\ver_{\TS}(\rho_n) = \ver_{\mathscr{M}}(\sigma)$.
  Moreover, obviously $\rho_0 \xrightarrow[\TS]{} \cdots \xrightarrow[\TS]{}
  \rho_n$ is complete run and thus we have found the desired
  validation for $\sigma$.
  
  For the converse direction, let $\TS$ be a test case that is valid
  w.\,r.\,t.\ $\mathscr{M}$. We need to show that the call {\bit
    valid}$(\mathscr{M},\TS, \epsilon)$ does not fail. Hence assume
  that is does fail and let $\rho \in L_{\mathscr{M}}$ a
  prefix-maximal word such that {\bit valid}$(\mathscr{M},\TS,\rho)$
  is evaluated. By definition of {\bit valid}, $\rho in L_\MS$ must
  holds.  One of the following choices for $\rho$ is the one that
  leads to failure.
  \begin{itemize}
  \item $\TS(\rho)$ is undefined, then obviously, for every $\sigma
    \in L_{\MS}$ with $\rho \leq \sigma$, $[\sigma]_{L_\MS} \cap
    L_{\TS} = \emptyset$.
  \item  $\TS(\rho) \in \FinVerdicts$ and $\TS(\rho)
    \neq \ver_{\mathscr{M}}(\rho)$, which violates the first condition
    in the definition of test case validity.
  \item $\TS(\rho) = \delta$ and {\bit en}$(\MS,\rho) \cap \ObsRec =
    \emptyset$. If $\ver_{\mathscr{M}}(\rho) \neq \Vnone$ then $L_{\TS}$
    and $L_{\mathscr{M}}$, then again the first condition of the
    definition of test case validity is violated. If
    $\ver_{\mathscr{M}}(\rho) = \Vnone$ then there exists $\sigma \in
    L_{\mathscr{M}}$ with $\ver_{\mathscr{M}}(\sigma) \neq \Vnone$ and
    $\rho < \sigma$. It is easy to see that, if there exists $\eta \in
    [\sigma]_{L_{\mathscr{M}}} \cap L_{\TS}$, then $\rho <
    \eta$, but since $\TS(\rho) = \delta$ and {\bit en}$(\MS,\rho) \cap
    \ObsRec = \emptyset$, for every $\chi \in L_{\TS}$ with $\rho < \chi$,
    $\chi \not \in L_{\mathscr{M}}$ holds and hence the does not
    exists such a $\eta$.
  \item The case that $\TS(\rho) = a \in \ObsSend$ but $\rho
    \cdot a \not \in L_{\mathscr{M}}$ is analog to the previous case.
  \end{itemize}
  
  Hence, {\bit valid}$(\mathscr{M},\TS,\epsilon)$ cannot fail.
\end{proof}

\begin{proof}[Proof of Theorem~\ref{theo:valid-tester-exists}]
  We define a test case $\TS$ as follows: for $\sigma \in
  \obs(X_{\mathscr{M}})$:
  \begin{gather*}
  \TS(\sigma) \df 
  \left\{\begin{array}{l} 
    \ver_{\mathscr{M}}(\delin(\mathbit{x})), 
      \text{if {\bit en}$(\MS,\sigma) = \emptyset$};\\
    \delta, \text{if $\emptyset \neq \text{\bit en}(\MS,\sigma) \subseteq \ObsRec$}\\
    a, \text{ for an arbitrary $a \in \text{\bit en}(\MS,\sigma) \cap \ObsSend$ otherwise}
  \end{array}\right.
  \end{gather*}
  
  It is easy to see that for this test case $\TS$, {\bit valid}$(\MS,\TS,\epsilon)$
  does indeed not fail and hence, by
  Theorem~\ref{theorem:alg-correct}, $\TS$ is a valid test case w.\,r.\,t.\ 
  $\MS$.
\end{proof}

\bibliographystyle{abbrv}
\bibliography{viva2}

\end{document}